\documentclass[modern]{aastex61}
\usepackage{amsmath}
\usepackage{bm}
\usepackage{graphicx}
\usepackage[caption=false]{subfig}

\newcommand{\boldPsi}{\boldsymbol{\Psi}}
\newcommand{\boldpsi}{\boldsymbol{\psi}}
\newcommand{\boldphi}{\boldsymbol{\phi}}

\newcommand{\acronym}[1]{{\small{#1}}}

\received{Month X, Year}
\revised{Month X, Year}
\accepted{Month X, Year}
\submitjournal{ApJ}
\shorttitle{Inferring binary and trinary stellar populations}
\shortauthors{Widmark et al.}
\begin{document}

\title{Inferring binary and trinary stellar populations in photometric and astrometric surveys}

\correspondingauthor{Axel Widmark}
\email{axel.widmark@fysik.su.se}

\author[0000-0001-5686-3743]{Axel Widmark}
\affiliation{Department of Physics, AlbaNova, Stockholm University, SE-106 91, Stockholm, Sweden}
\affiliation{The Oskar Klein Centre for Cosmoparticle Physics, AlbaNova, SE-106 91 Stockholm, Sweden}

\author[0000-0002-3962-9274]{Boris Leistedt}
\affiliation{Center for Cosmology and Particle Physics, Department of Physics,  \\ \quad New York University, 726 Broadway, New York, NY 10003, USA}
\affiliation{NASA Einstein Fellow}

\author[0000-0003-2866-9403]{David W. Hogg}
\affiliation{Center for Cosmology and Particle Physics, Department of Physics,  \\ \quad New York University, 726 Broadway, New York, NY 10003, USA}
\affiliation{Center for Data Science, New York University, 60 Fifth Avenue, New York, NY 10011, USA}
\affiliation{Flatiron Institute, 162 Fifth Avenue, New York, NY 10010, USA}



\begin{abstract}
Multiple stellar systems are ubiquitous in the Milky Way, but are often unresolved and seen as single objects in spectroscopic, photometric, and astrometric surveys.
Yet, modeling them is essential for developing a full understanding of large surveys such as Gaia, and connecting them to stellar and Galactic models. In this paper we address this problem by jointly fitting the Gaia and 2MASS photometric and astrometric data using a data-driven Bayesian hierarchical model that includes populations of binary and trinary systems. This allows us to classify observations into singles, binaries, and trinaries, in a robust and efficient manner, without resorting to external models. We are able to identify multiple systems and, in some cases, make strong predictions for the properties of its unresolved stars. We will be able to compare such predictions with Gaia Data Release 4, which will contain astrometric identification and analysis of binary systems.
\end{abstract}



\section{Introduction}
A large fraction of stars, possibly a majority, exist in binaries and higher multiple stellar systems \citep{Moe17,Duchene13}. In stellar surveys, such systems are resolved only if they are sufficiently close. When unresolved, they are reported as single objects where the respective stellar fluxes are added into a total magnitude. Even though a multiple stellar system cannot be resolved by pure angular separation, there are other techniques to do so. With multi-epoch photometry it can be possible to observe eclipses, depending on the spatial orientation of the system. For very tight binaries with high rotational periods, spectroscopy with high time resolution can be used to infer the rotation from a periodic widening/narrowing of stellar emission lines, as applied to white dwarf binaries in the Supernova Progenitor Survey \citep[SPY,][]{Maoz16}. Also, without such time variations, clues are given by spectral or photometric information. For example, the sum of two black body spectra of different temperatures is not itself a black body spectrum. With good stellar models it is possible to disentangle such systems, as discussed in \cite{ElBadry17}. In this article, we use a data-driven model to infer multiple stellar systems, meaning that we do not rely on stellar models. The strength of this approach is its generality, in that we fit our model to the data with minimal a priori assumptions about the stellar population. In the era of large stellar surveys such as Gaia \citep{gaia}, it should be possible to simultaneously model and fit for single and multiple systems without the need of external models, as we demonstrate in the present work. While other identification methods are observationally expensive and often require special purpose surveys, this method relies on readily available photometry.

In this paper we focus on the identifying multiple stellar systems in the first data release of the Gaia survey \citep{gaia_dr1}. The aim is to infer what stellar objects in the Gaia catalogue are binary or trinary from two-band photometry and an astrometric distance measurement. Combining two stars from the main sequence into an unresolved binary does not result in a stellar system that is itself consistent with the main sequence distribution, at least not where the main sequence is sufficiently narrow in color-magnitude space. For example, combining two identical stars will double the flux, giving identical color but a magnitude difference of $\Delta M \simeq 0.75$.

We use a simple hierarchical Bayesian model in a small window of color-magnitude space, with cuts on observed parallax (giving a volume limited sample). We assume a main sequence of single stellar systems that is Gaussian in width and infer a population of binary and trinary stellar systems. This is a powerful way to identify multiple stellar systems and in some cases even infer the color and magnitude of the individual stars, which in turn could be turned into constraints on the individual masses and temperatures using stellar models.

In the longer term, we hope to include this formalism in a larger framework of data-driven models, as used in \cite{Leistedt17,Anderson17}. In both these sources, the authors use Gaussian Mixture Models. With this long term goal in mind, we describe the population of stellar systems in color-magnitude space in terms of a mixture of multivariate Gaussians.

By the end of its mission, Gaia is predicted to have photometry errors in the order of milli-magnitudes. However, for now the photometry is nowhere near that level of precision, and does not provide color information, which is why we in this article use photometric information from the 2MASS survey.

This article is outlined as follows. In section \ref{sec:catalogue}, we describe the catalogue and the sample of data we use. In section \ref{sec:model}, we present the statistical model for the stellar population. In section \ref{sec:disentangle}, we show how to infer properties of component stars in a multiple stellar system. In section \ref{sec:results}, we present our results. In section \ref{sec:discussion}, we discuss and conclude.

\section{TGAS/2MASS cross match}\label{sec:catalogue}

In this work we use a catalogue of astrometric data from the Tycho-2 Gaia Astrometric Solution \citep[TGAS,][]{tgas} cross matched with photometric data from the Two Micron All Sky Survey (2MASS). The cross match is taken from \cite{Smart16}.

\subsection{Observables and errors}\label{sec:observables}

The observables we take into account are parallax, $\varpi$, and apparent magnitudes in $J$ and $K$ bands, $m_J$ and $m_K$. A hatted quantity ($\hat{\varpi}$) refers to the observed value, while a non-hatted quantity ($\varpi$) refers to its true value. We operate in the space of color and absolute magnitude in the $J$-band, where the true and observed quantities are

\begin{equation}
\begin{split}
	c & = m_J-m_K, \\
    \hat{c} & = \hat{m}_J-\hat{m}_K, \\
    M & = m_J - 5\Big[\log_{10}\Big(\frac{\text{arcs}}{\varpi}\Big)-1\Big], \\
    \hat{M} & = \hat{m}_J - 5\Big[\log_{10}\Big(\frac{\text{arcs}}{\hat{\varpi}}\Big)-1\Big].
\end{split}
\end{equation}
In this article, the only absolute magnitude considered is that of the $J$-band, so for the sake of brevity we write the absolute magnitude without the $J$-index.

We propagate the error on the parallax to first order, giving a error covariance matrix over $c$ and $M$

\begin{equation}
    \Sigma_{c,M}^{(i)} = 
    \begin{bmatrix}
        \sigma_J^2+\sigma_K^2 & & \sigma_J^2 \\
        \sigma_J^2 & & \sigma_J^2 + (\frac{5\sigma_\varpi}{\hat{\varpi}\log 10})^2
    \end{bmatrix},
\end{equation}
with an implicit index $i$ that denotes the stellar object. This first order propagation is motivated by a high parallax significance, $\hat{\varpi}/\sigma_\varpi$. Objects in our sample have a median parallax significance of 25 and a minimum value of 5. See section \ref{sec:cuts} below for details on how the sample is constructed.

Because the population is modeled as a Gaussian mixture, it is very convenient to have errors also in the form of a covariance matrix. Computing the posterior becomes more efficient, since the convolution of two bivariate Gaussians has a simple analytical form.

We do not consider dust extinction corrections to the photometric magnitudes, as these effects are expected to be negligible in the color--magnitude box we consider given our volume limited sample. To check this, we calculated the dust corrections using the \cite{Green2015bayestar} three-dimensional dust map, evaluated at the point distance estimate $1/\hat{\varpi}$ for each object. We find that the median dust corrections to the $J$ and $K$ band magnitudes are 0.0068 and 0.0029, respectively, while the 90th percentiles are 0.0412 and 0.0174. These corrections are small compared to the width of the components of our Gaussian mixtures, and we neglect them in order to simplify our analysis.

\subsection{Data cuts}\label{sec:cuts}

We make cuts to the data according to the following criteria:

\begin{itemize}
	\item an observed parallax $\hat{\varpi}>5$ mas (corresponding to a distance $<200$ pc),
    \item an observed color $\hat{c}$ in interval 0.5--0.8,
    \item an observed absolute magnitude $\hat{M}$ in interval 3--6,
    \item no NaN-valued observables,
    \item no excessive photometric noise, defined as $\sigma_J^2+\sigma_K^2<0.1^2$ (this criterion removes 11 objects from our sample).
\end{itemize}

\begin{figure}[tbp]
\centering
\includegraphics[width=1.\textwidth,origin=c]{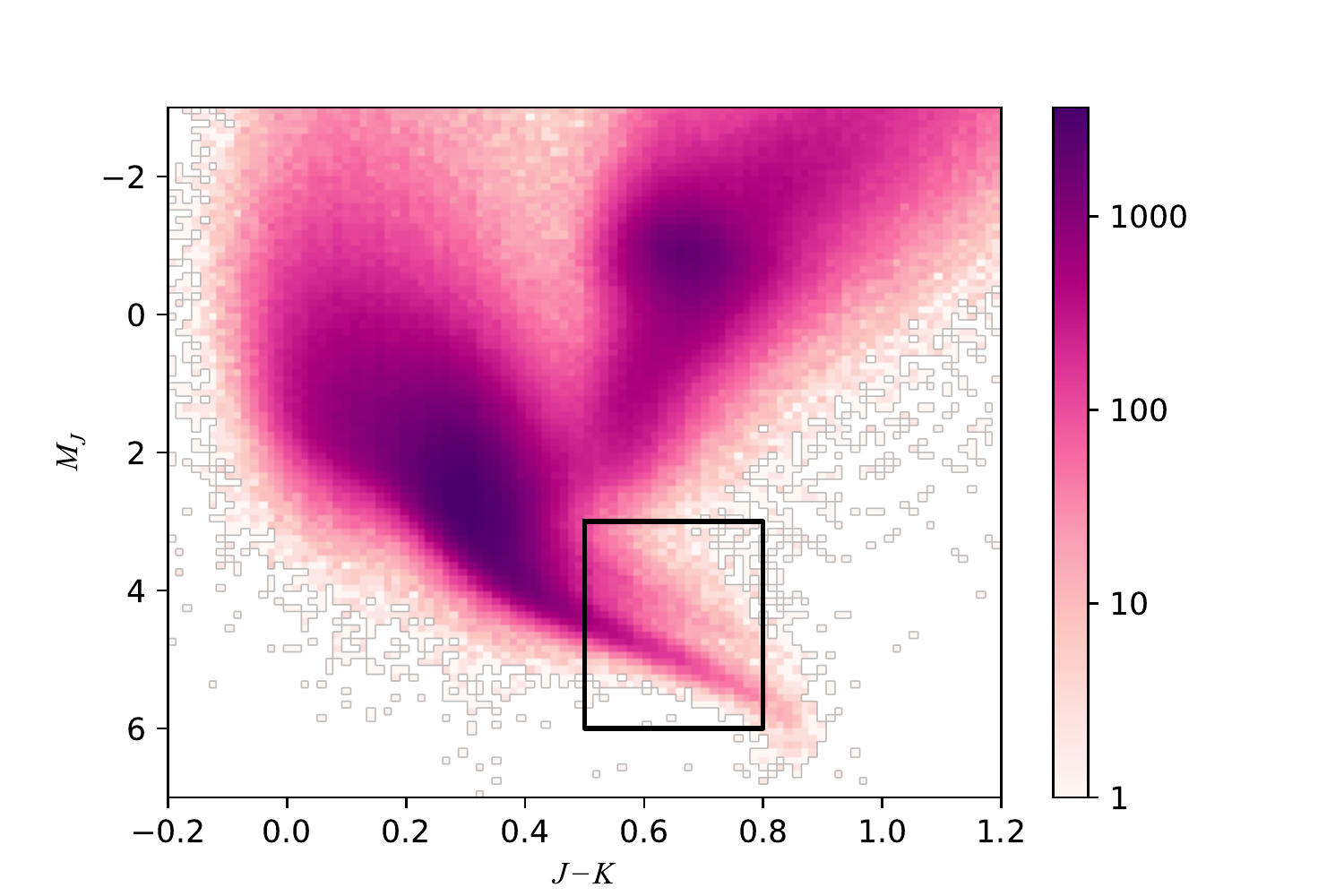}
\caption{\label{fig:HR_cut}A 2d-histogram of the TGAS/2MASS data. The cuts in color and magnitude by which we construct our sample are illustrated by the black square. In this subregion, a possible binary sequence is visible above the main sequence, as a second band of slightly brighter objects. The color corresponds to the number of stars per bin, where the scale is logarithmic and the number of bins is $100 \times 100$. No dust corrections have been applied to the data in this figure (see text for details).}
\end{figure}

This gives a total of 23,112 individual objects. The full color-magnitude diagram of the TGAS/2MASS cross-match is shown in figure \ref{fig:HR_cut}, also marking the smaller window of parameter space that we operate in. These cuts on color and magnitude limits our single star population to K-type dwarfs \citep{Pecaut13}.

\section{Model}\label{sec:model}

In this section we outline the hierarchical model that includes binaries and trinaries in our window of color-magnitude space. The distribution of single stars is modeled by Gaussians generated from population parameters $\alpha$, $\beta_1$, $\beta_2$, and $\gamma$. These Gaussians are centered on a straight line, according to

\begin{equation}\label{eq:mu}
    M = \beta_1 + \beta_2 (c-0.5),
\end{equation}
with a spacing of 0.05 in color between Gaussians. The covariance matrix describing the Gaussians have one eigenvector pointing along the line described by equation \eqref{eq:mu}, which is set to width 0.05 in $c$, and one eigenvector perpendicular to said line of width $\gamma$.

The amplitude, \textit{i.e.} the number of stars in each Gaussian, is proportional to

\begin{equation}\label{eq:numberdensity}
    n_\text{singles}(c) \propto e^{-\alpha c}.
\end{equation}

In other words, the parameter $\alpha$ describes how the number density changes with color; $\beta_1$ and $\beta_2$ describe the slope of the distribution; $\gamma$ describes the thickness of the main sequence distribution of single stars. The Gaussian mixture that describe the population of single stars has parameters

\begin{equation}
\begin{split}
	\bar{c}_s &= 0.05s \\
    \bar{M}_s &= \beta_1 + \beta_2(0.05s-0.5) \\
    \Sigma_{(s)} &= \begin{bmatrix}
        0.05^2+\beta_2^2(\beta_2^2+1)^{-1}\gamma^2
        & \quad & 0.05^2\beta_2+\beta_2(\beta_2^2+1)^{-1}\gamma^2\\
        0.05^2\beta_2+\beta_2(\beta_2^2+1)^{-1}\gamma^2
        & \quad & 0.05^2\beta_2^2 + (\beta_2^2+1)^{-1}\gamma^2
    \end{bmatrix},
\end{split}
\end{equation}
where $s$ is the index of the Gaussian and the bar in $\bar{c}$ and $\bar{M}$ indicate that they are mean values of the respective Gaussians. The integer on the covariance matrix $\Sigma_{(s)}$ is in parenthesis, as it is identical for all Gaussians of the singles population. The two eigenvectors are manifest as the two separate terms of the respective matrix element. Because we limit our sample to an interval in color, $c\in[0.5,0.8]$, the Gaussian mixture of the singles population includes Gaussian with a mean color in region $\bar{c}\in[0.4,0.9]$, giving a total of 11 Gaussians.
Despite the simplicity of this model, we expect from stellar models and existing data that it will provide an accurate description of the population of single stars.

In addition to aforementioned four population parameters, there are three more population parameters that describe the binary and trinary populations: $\eta$, which parameterizes the pairing mechanism of multiple stellar systems (how favored are systems where the component stars have similar properties); and the fraction of binary and trinary systems, $f_{b}$ and $f_{t}$, corresponding to the (relative) prior probabilities of an object being an unresolved binary or trinary system as opposed to a single star.

\subsection{Gaussian mixture and binary population}

We modeled the population of single star objects as a sum of Gaussians in color-magnitude space. The population of binary and trinary stellar systems are sums of the single population Gaussians. Adding up the magnitude of the objects making an unresolved multiple system follows the formula

\begin{equation}\label{eq:addmags}
    M_\text{sum} = -\frac{5}{2}\log_{10}\left( \sum_i 10^{ -\frac{2}{5}M_i } \right).
\end{equation}
Because this is not a straight forward addition, the sum of two Gaussian distributions in color-magnitude space is not exactly a Gaussian itself. However, the actual result is close to a Gaussian distribution and can be approximated as such to an adequate degree of accuracy. Given two Gaussians with means $\{\bar{c}_i,\bar{M}_i\}$ and covariance matrices $\Sigma_i$, where $i=\{1,2\}$, a first order approximation to the sum is a Gaussian distribution, where the means are added together according to equation \eqref{eq:addmags}, and the summed covariance matrix is given by

\begin{equation}\label{eq:analyticgaussiansum}
    \Sigma_\text{sum} = \mathcal{J}_1 \Sigma_1 \mathcal{J}_1^\top +  \mathcal{J}_2 \Sigma_2 \mathcal{J}_2^\top.
\end{equation}
Here, $\mathcal{J}_{i=\{1,2\}}$ are Jacobians of the form

\begin{equation}\label{eq:jacobians}
\begin{split}
    \mathcal{J}_i & = \begin{bmatrix}
        \frac{\displaystyle \partial c_\text{sum}}{\displaystyle \partial c_i}\Big|_{c_i=\bar{c}_i}
        & & \frac{\displaystyle \partial c_\text{sum}}{\displaystyle \partial M_i}\Big|_{M_i=\bar{M}_i} \\[2ex]
        \frac{\displaystyle \partial M_\text{sum}}{\displaystyle \partial c_i}\Big|_{c_i=\bar{c}_i}
        & & \frac{\displaystyle \partial M_\text{sum}}{\displaystyle \partial M_i}\Big|_{M_i=\bar{M}_i}
    \end{bmatrix} = \\
    & = \begin{bmatrix}
        \frac{\displaystyle g(\bar{M}_i-\bar{c}_i)}{\displaystyle g(\bar{M}_1-\bar{c}_1)+g(\bar{M}_2-\bar{c}_2)}
        & & \frac{\displaystyle g(\bar{M}_i)}{\displaystyle g(\bar{M}_1)+g(\bar{M}_2)} - \frac{\displaystyle g(\bar{M}_i-\bar{c}_i)}{\displaystyle g(\bar{M}_1-\bar{c}_1)+g(\bar{M}_2-\bar{c}_2)} \\[2ex]
        0
        & & \frac{\displaystyle g(\bar{M}_i)}{\displaystyle g(\bar{M}_1)+g(\bar{M}_2)}
    \end{bmatrix},
\end{split}
\end{equation}
written in terms of the function $g(x)=10^{-2x/5}$.

\begin{figure}[tbp]
\centering
\includegraphics[width=.6\textwidth,origin=c]{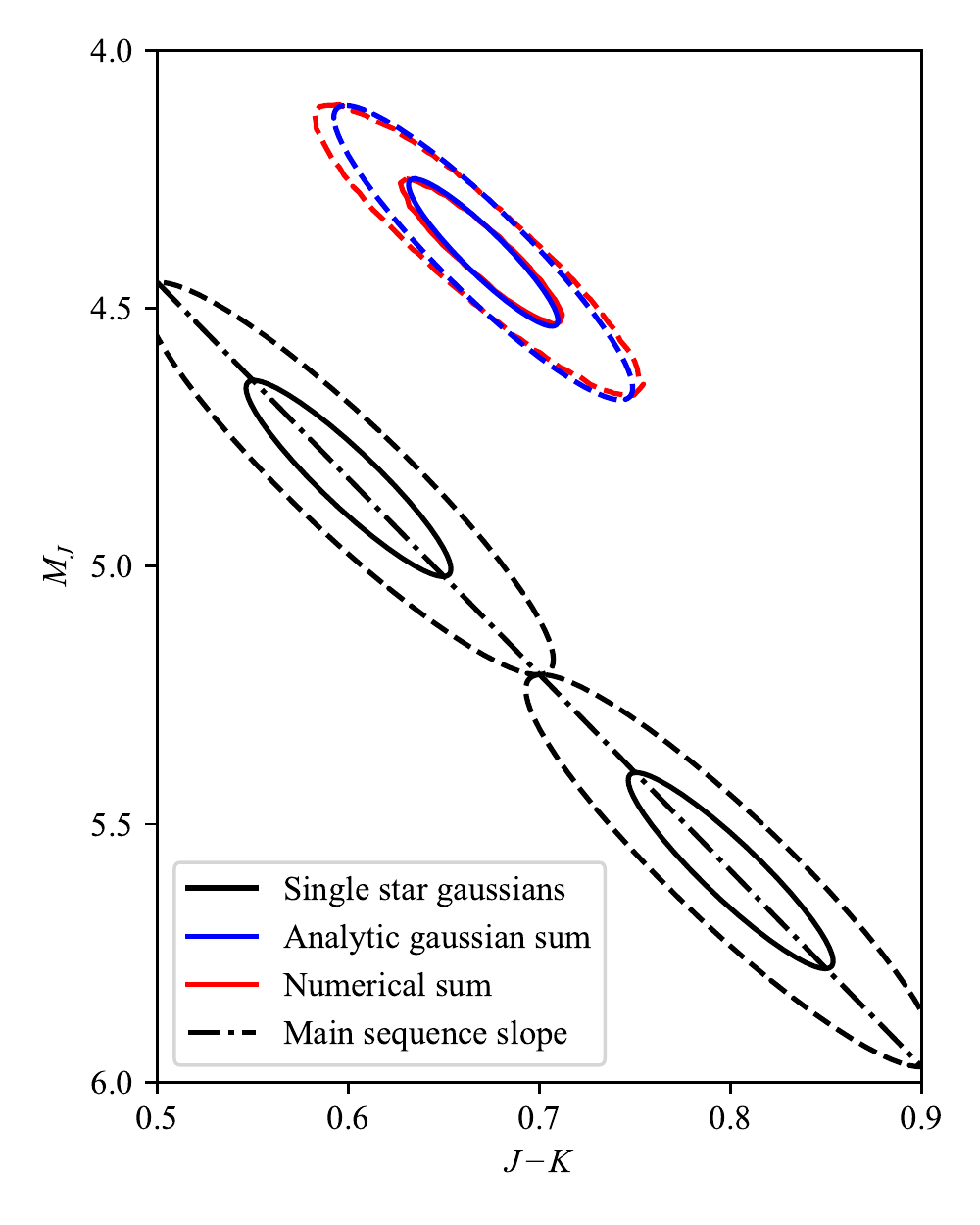}
\caption{\label{fig:gaussapprox}Example of the analytic addition of single stellar Gaussian distributions. The black dash-dotted line is the main sequence slope, assuming parameter values $\beta_1=4.45$ and $\beta_2=3.8$. In black are two Gaussian distributions representing single stellar systems, assuming parameter values $\beta_1=4.45$, $\beta_2=3.8$, and $\gamma=0.02$, where the solid (dashed) line represents its $1\sigma$ ($2\sigma$) regions. In blue we see the same for the analytically derived binary bivariate distribution, as a sum of the two single distribution according to equation \eqref{eq:analyticgaussiansum}. In red is the numerical solution, found by adding randomly drawn single stars together. This converges to the exact solution but is computationally expensive.}
\end{figure}

While this is only propagating the Gaussians to first order, it does work very well, as is evident from figure \ref{fig:gaussapprox}. The approximation works well for adding Gaussians that are sufficiently small, such that higher order terms does not come into significant effect.

For the Gaussian mixtures of the binary and trinary populations, the amplitude of the Gaussian is proportional to $\exp(-\alpha \bar{c})\exp(\eta \Delta_{\bar{c}})$, where $\Delta_{\bar{c}}=|\bar{c}_A-\bar{c}_B|$ for binaries and $\Delta_{\bar{c}}=|\bar{c}_A-\bar{c}_C|$ for trinaries, which is the difference in color between the highest and lowest Gaussian mean color values. The first factor of the amplitude is directly proportional to the number density of single stars of the same color. The seconds factor comes from the fact that multiple stellar systems with similar properties might be favored or disfavored, which we parameterize with population parameter $\eta$. We would expect that binaries with stars of approximately equal mass, and similar color, are more likely than binaries with a significant mass difference \citep{Moe17,Duchene13}.

As in the case of the Gaussian mixture that describes the singles population, we constrain ourselves to include Gaussians that have a mean color in the interval $\bar{c}\in[0.4,0.9]$. A Gaussian in the mixture that describes the binary or trinary population is built from two or three parent Gaussians belonging to the singles population. The parent Gaussians do not necessarily have a mean color within the interval $\bar{c}\in[0.4,0.9]$, even though the composite stellar object does so. For example, it is possible to have a very dim component star, of large color value $c$, such that its contribution to the total color is small. For this reason we assume the distribution of singles to be extended analytically beyond this window according to the model. As the hydrogen burning limit is reached around $c=1.1$ \citep{Pecaut13}, the outermost single star parent Gaussian has mean color $\bar{c}=1.05$. This makes the distribution of single stars fall off smoothly at $c=1.1$.

The total number of Gaussians in the binary Gaussian mixtures is 89. For the case of the trinary population, the number of Gaussians is very large, and unnecessarily so. Therefore we do not use all possible combinations of parent stars. Rather, the three parents are given by three single stars with indices $(s,s+2k,s+2l)$, where $s,k,l$ are integers fulfilling that $l\geq k\geq 0$. This gives a total of 193 Gaussians for the trinary Gaussian mixture. Although we are thinning the number of trinary Gaussians, the total number of trinary stellar systems are unaffected; this is accounted for by a normalization factor, as we shall see in section \ref{sec:hierarchicalmodel}.

\subsection{Hierarchical model}\label{sec:hierarchicalmodel}

There is a total of 7 population parameters, capsuled in a vector $\boldPsi$, as is listed in table \ref{tab:parameters}. Also visible in the table are the parameters of the Gaussian mixture (color and magnitude mean values and a $2 \times 2$ covariance matrix), the stellar object parameters (intrinsic color and magnitude), and the data of an observed object (observed color and magnitude and a $2 \times 2$ covariance error matrix). When writing the object parameters or data without an index $i$, we refer to the stellar parameters or data of all objects, like $\boldpsi = \{ \boldpsi_1,\boldpsi_2,...,\boldpsi_N \}$. The index $i$ always denotes a stellar object, while the index $s$, $b$, and $t$ always denotes a Gaussian in the Gaussian mixture of the single, binary, or trinary population.

\begin{table}
	\centering
	\caption{List of the 7 population parameters, the parameters of the Gaussian mixture, the intrinsic stellar object parameters, and the data of an observed object.}
	\label{tab:parameters}
    \begin{tabular}{l}
		\hline
		Population parameters, $\boldPsi$ \\
		$\alpha$,\quad $\beta_1$,\quad $\beta_2$,\quad $\gamma$,\quad $\eta$,\quad $f_{b}$,\quad $f_{t}$ \\
        \hline
		Parameters of the $s$th Gaussian \\
		$\bar{\boldphi}_s = \{ \bar{c}_s,\bar{M}_s \}$,\quad $\Sigma_s$ \\
		\hline
		Stellar object parameters, $\boldpsi_{i=1,...,n}$ \\
        $c = m_J-m_K$,\quad $M$ \\
        \hline
        Data (observables and errors), $\mathbf{d}_{i=1,...,n}$ \\
        $\hat{\boldphi}_i = \{\hat{c}_i,\hat{M}_i\}$,\quad $\Sigma_i$ \\
        \hline
	\end{tabular}
\end{table}

The posterior on the population parameters $\boldPsi$ and stellar parameters $\boldpsi$ is written

\begin{equation}
    \text{Pr}(\boldPsi,\boldpsi|\mathbf{d}) = 
    \text{Pr}(\boldPsi) \prod_i \text{Pr}(\mathbf{d}_i | \boldpsi_i)\text{Pr}(\boldpsi_i | \boldPsi).
\end{equation}
The prior demands that $\gamma$, $f_{b}$, and $f_{t}$ are positive and that $f_{b}+f_{t}<1$, and is otherwise flat in all parameters, like

\begin{equation}
\text{Pr}(\boldPsi) = \theta(f_{b})\, \theta(f_{t})\, \theta(1-f_{b}-f_{t})\, \text{Pr}(\alpha, \beta_1, \beta_2, \gamma, \eta),
\end{equation}
where $\theta$ is the Heaviside step function and $\text{Pr}(\alpha, \beta_1, \beta_2, \gamma, \eta)$ is uniform in some arbitrarily large volume in parameter space.

The posterior density of the population parameters, with stellar parameters marginalized, is

\begin{equation}
    \text{Pr}(\boldPsi|\mathbf{d}) = 
    \text{Pr}(\boldPsi) \prod_i \int \text{d}\boldpsi_i \text{Pr}(\mathbf{d}_i | \boldpsi_i)\text{Pr}(\boldpsi_i | \boldPsi).
\end{equation}
Because both the errors are Gaussian and the model is a sum of Gaussians, the integral over an individual object can be written as a sum over bivariate Gaussian distributions, like

\begin{equation}\label{eq:1objectposterior}
\begin{split}
    & \int \text{d}\boldpsi_i \text{Pr}(\mathbf{d}_i | \boldpsi_i)\text{Pr}(\boldpsi_i | \boldPsi) =  \\
    \frac{1-f_{b}-f_{t}}{\bar{N}_\text{single}} & \sum_{\substack{\text{single} \\ \text{Gaussians} \\ s}} \exp(-\alpha \bar{c}_s)\mathcal{N}(\hat{\boldphi}_i-\bar{\boldphi}_s,\Sigma_i+\Sigma_s)  \\
    + \frac{f_{b}}{\bar{N}_\text{binary}} & \sum_{\substack{\text{binary} \\ \text{Gaussians} \\ b}} \exp(-\alpha \bar{c}_b)\exp(\eta \Delta_{\bar{c},b})\mathcal{N}(\hat{\boldphi}_i-\bar{\boldphi}_b,\Sigma_i+\Sigma_b)  \\
    + \frac{f_{t}}{\bar{N}_\text{trinary}} & \sum_{\substack{\text{trinary} \\ \text{Gaussians} \\ t}} \exp(-\alpha \bar{c}_t)\exp(\eta \Delta_{\bar{c},t})\mathcal{N}(\hat{\boldphi}_i-\bar{\boldphi}_t,\Sigma_i+\Sigma_t).
\end{split}
\end{equation}
The mean color $\bar{c}_s$ and $\alpha$ sets the amplitude of the $s$th Gaussian of the single population, and for binaries (trinaries) there is the additional factor given by $\eta$ and $\Delta_{\bar{c},b}$ ($\Delta_{\bar{c},t}$). The quantities $\bar{N}_\text{single}$, $\bar{N}_\text{binary}$, $\bar{N}_\text{trinary}$ are normalizations equal to the integrated distribution for the population in our color-magnitude sample window. In the case of the population of single stars (and equivalently in the two other cases), we have that

\begin{equation}
\begin{split}
	\bar{N}_\text{single} = \sum_{\substack{\text{single} \\ \text{Gaussians} \\ s}}
    \exp(-\alpha \bar{c}_s)\frac{1}{4}
    \Bigg[\text{erf}\Bigg( \frac{0.8-\bar{c}_s}{\sqrt{2(\sigma_{c,s}^2+\tilde{\sigma}_c^2)}} \Bigg)
    +\text{erf}\Bigg( \frac{\bar{c}_s-0.5}{\sqrt{2(\sigma_{c,s}^2+\tilde{\sigma}_c^2)}} \Bigg) \Bigg] \\ \times
    \Bigg[\text{erf}\Bigg( \frac{6.0-\bar{M}_s}{\sqrt{2(\sigma_{M,s}^2+\tilde{\sigma}_M^2)}} \Bigg)
    +\text{erf}\Bigg( \frac{\bar{M}_s-3.0}{\sqrt{2(\sigma_{M,s}^2+\tilde{\sigma}_M^2)}} \Bigg) \Bigg],
\end{split}
\end{equation}
which is a sum over all Gaussians in the mixture, of their respective amplitudes times the probability that a star of that Gaussian will be included in the window of parameter space. The quantities $\sigma_{c,s}$ and $\sigma_{M,s}$ are the width of the $s$th Gaussian in color and magnitude, \textit{i.e.} the diagonal elements of $\Sigma_s$. The quantities $\tilde{\sigma}_c$ and $\tilde{\sigma}_M$ are the median errors to the color and magnitude data, accounting for the fact that observational errors can cause an object that is actually inside the allowed region of parameter space to be mistaken for being outside, and vice versa. For our sample, they have values $\tilde{\sigma}_c=0.032$ and $\tilde{\sigma}_M=0.090$. We have made some approximations here: we take the median values instead of integrating over the full distribution of errors (where this distribution could even be a function of $c$ and $M$); furthermore, we do not account for the non-diagonal elements of the covariance matrices. Both of these factors are small and have minuscule bearing on our end result. The simplification of taking the median error is only applied when calculating the normalization factor $\bar{N}$; the individual errors of each object is accounted for in all other factors of equation \eqref{eq:1objectposterior}.

\begin{figure}[tbp]
\centering
\includegraphics[width=.45\textwidth,origin=c]{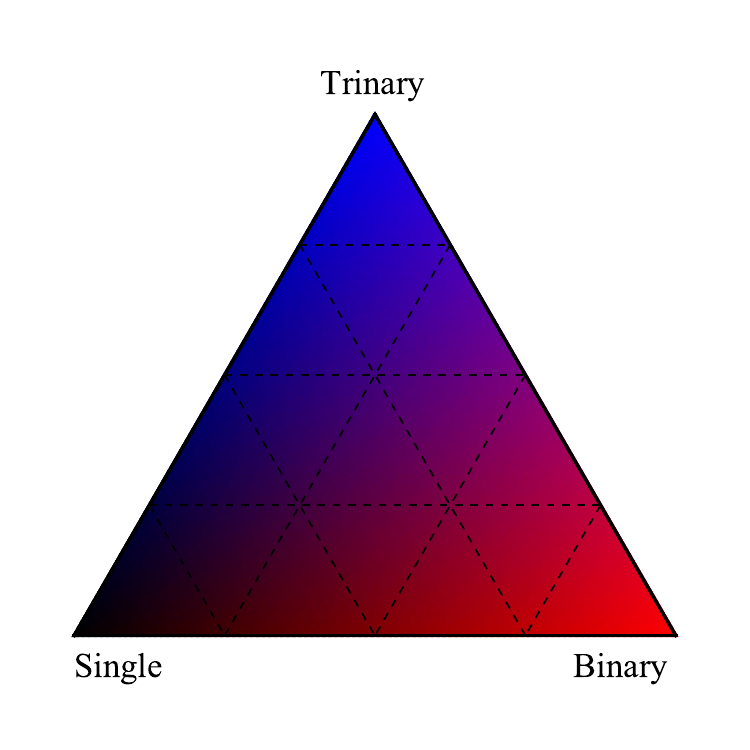}
\caption{\label{fig:colortriangle}Color code denoting the object type (single, binary, or trinary stellar system), used throughout this article. The color is continuous between the three object types. For example, an object that is equally likely of being single, binary, and trinary, will be located in the dark purple center of the triangle.}
\end{figure}

The probability of a stellar object $x$ being binary, given population parameters, is given by

\begin{equation}
	\text{Pr}( x \text{ is binary} | \mathbf{d}_x, \boldPsi) =
    \dfrac{\displaystyle \frac{f_b}{\bar{N}_\text{binary}}\sum_{b} \exp(-\alpha \bar{c}_b)\exp(\eta \Delta_{\bar{c},b})\mathcal{N}(\hat{\boldphi}_x-\bar{\boldphi}_b,\Sigma_x+\Sigma_b)}{\displaystyle \int \text{d}\boldpsi_x \text{Pr}(\mathbf{d}_x | \boldpsi_x)\text{Pr}(\boldpsi_x | \boldPsi)}.
\end{equation}
The denominator is a normalization as given by equation \eqref{eq:1objectposterior}. The probability of an object being single or trinary is expressed in an analogous manner. The probability of being single, binary and trinary adds up to unity and can be expressed using a color code according to figure \ref{fig:colortriangle}.

\section{Disentangling binary and trinary systems}\label{sec:disentangle}
Given our model, where the main sequence is narrow and Gaussian in width, it is not only possible to infer that a stellar object is a binary or a trinary system; one can also constrain the properties of the component stars. In this section we only make the binary case explicit, but analogous equations can be straightforwardly derived for the trinary case.

Given population parameters $\boldPsi$ and assuming that the stellar object with data $\mathbf{d}_x = \{\hat{\boldphi}_x,\Sigma_x\}$ is indeed a binary, the posterior on the component stars' properties, $\boldpsi_{A}=\{c_{A},M_{A}\}$ and $\boldpsi_{B}=\{c_{B},M_{B}\}$, is proportional to

\begin{equation}
	\text{Pr}(\boldpsi_A,\boldpsi_B|\boldPsi,\mathbf{d}_x) \propto
    \mathcal{N}(\hat{\boldphi}_x-\boldpsi_{A+B},\Sigma_x) \text{Pr}(\boldpsi_A,\boldpsi_B | \boldPsi)
\end{equation}
where $\boldpsi_{A+B}$ is the color and absolute magnitude of single stars $A$ and $B$ combined, following equation \eqref{eq:addmags}.

The population of binaries is modeled with a Gaussian mixture, where each Gaussian has two parent single star Gaussians. However, to get the probability density of the two component stars, $\text{Pr}(\boldpsi_A,\boldpsi_B | \boldPsi)$, we have to look at individual stars. The number density of binaries decays with the color of the binary (not the colors of its component stars), according to the same exponential law as the single star population. The probability density of the two component stars becomes

\begin{equation}
\begin{split}
\text{Pr}(\boldpsi_A,\boldpsi_B | \boldPsi) \propto \\
    \exp(-\alpha c_{A+B})
    \exp(\eta |c_A-c_B|)
    \sum_{s}
    \mathcal{N}(\boldpsi_A-\bar{\boldphi}_s,\Sigma_s)
    \sum_{s'}
    \mathcal{N}(\boldpsi_B-\bar{\boldphi}_{s'},\Sigma_{s'})
\end{split}
\end{equation}

When inferring the properties of individual objects, we do so in the regime of Empirical Bayes, in the sense that we reuse the information on the inferred population parameters as if it was independent of the object in question\footnote{\url{https://en.wikipedia.org/wiki/Empirical_Bayes_method}}. This is formally incorrect but since we are working with a sample of so many objects, removing one object from the sample has a negligible effect on the population parameter posterior.

\section{Results}\label{sec:results}

\begin{figure}[tbp]
\centering
\includegraphics[width=\textwidth,origin=c]{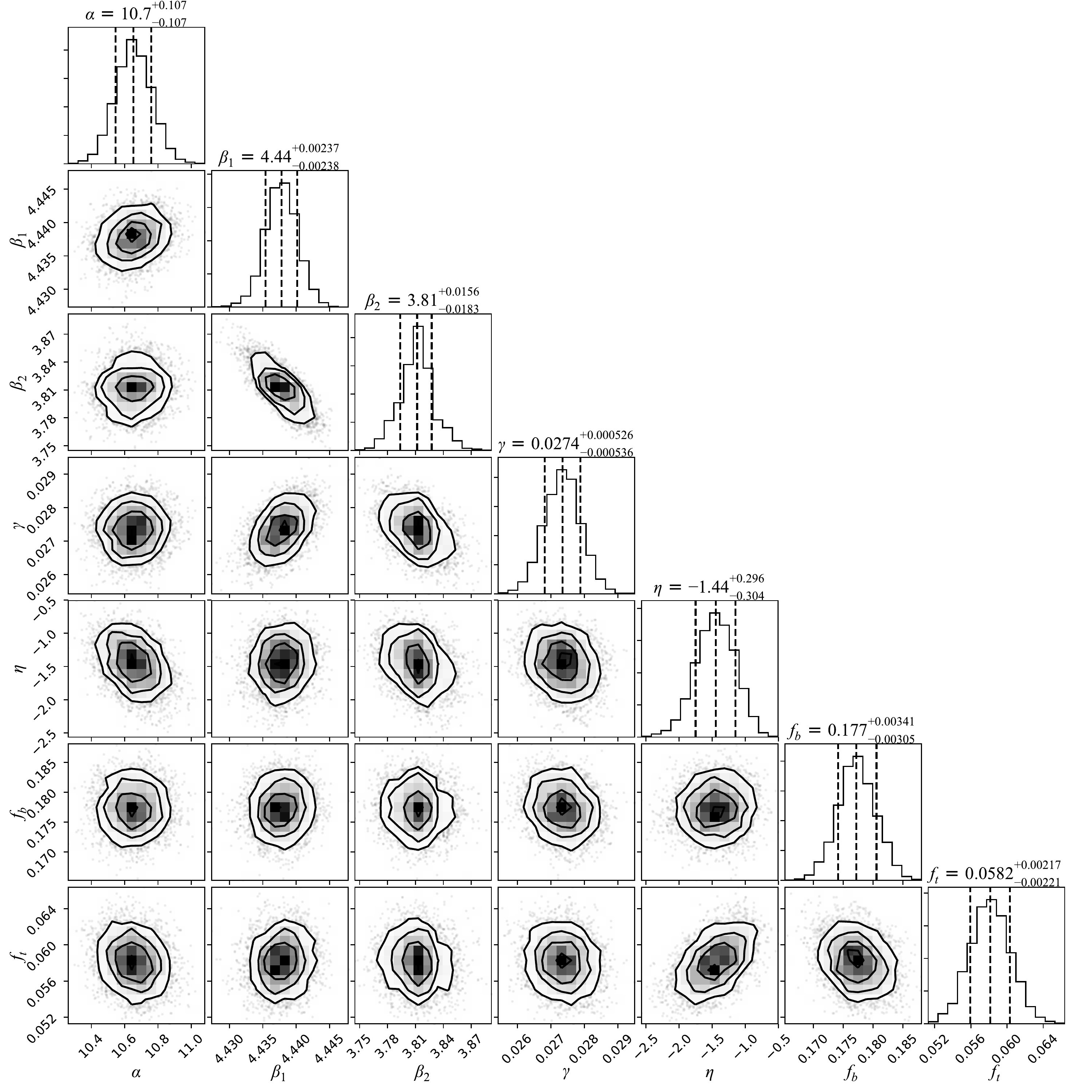}
\caption{\label{fig:posterior}Posterior of the 7 population parameters, as summarized in table \ref{tab:parameters}. The values presented on top, as well as the dotted lines in the 1-d histograms, are presented in terms of the median values plus/minus the 16th and 84th percentiles.}
\end{figure}

The posterior density for the 7 population parameters is visible in figure \ref{fig:posterior}. Those were obtained by histograming samples obtained with a Monte-Carlo Markov Chain, using a Metropolis random walk of 40,000 steps, thinned by a factor 10. The step length was calibrated in a thorough burn-in phase.

The posterior density is well constrained. The value for $\alpha$ is largely due to selection effects, as the true abundance of stars does not decrease with color. The other parameters of the main sequence slope ($\beta_1$, $\beta_2$, and $\gamma$) and have reasonable values and are strongly constrained. The parameters of binary and trinary populations ($\eta$, $f_b$, and $f_t$) also well constrained, although these values are heavily influenced by selection effects.

\begin{figure}[tbp]
\centering
\includegraphics[width=\textwidth,origin=c]{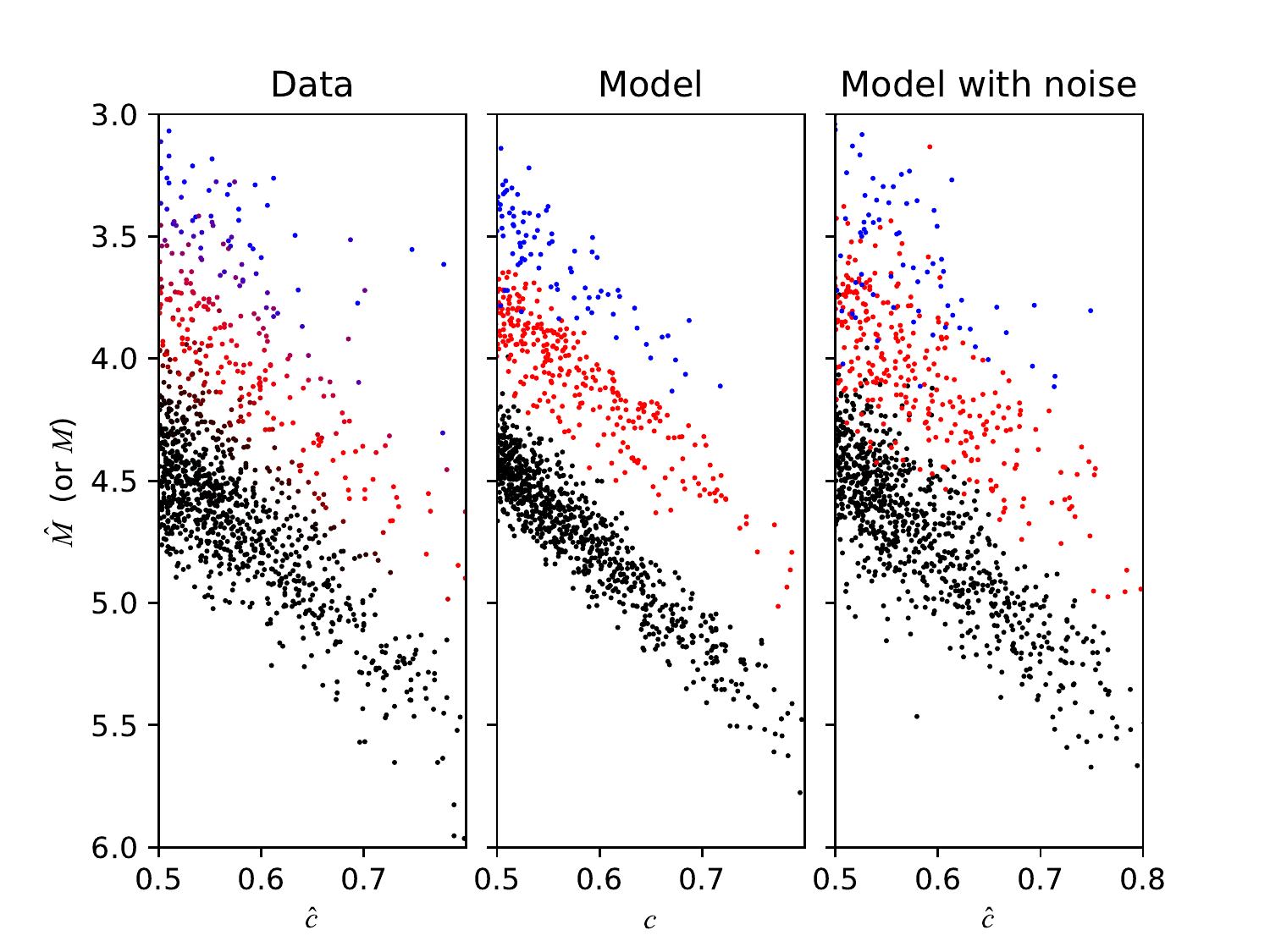}
\caption{\label{fig:data&model}Scatter plot for the data points, the model, and the model with noise. For visibility, only every 20th object of our sample is shown. The color to the data points (left panel) correspond to the relative posterior probabilities of being single/binary/trinary, marginalized over the population parameters, following the color code as presented in figure \ref{fig:colortriangle}. In the panels of the model, the type is known, so the colors are not continuous but black, red, or blue. The model is plotted for median posterior values to the population parameter posterior.}
\end{figure}

In figure \ref{fig:data&model} we see a scatter plot of a thinned set of data points, the highest posterior model without noise, and the model with noise. In the model panels, the black/red/blue correspond to single/binary/trinary systems. In the data panel, the color corresponds to the posterior probability of the object type, marginalized over the population parameters. The color scale is continuous according to figure \ref{fig:colortriangle}, so some points are for example purple, signifying that binary and trinary classifications are roughly equally likely. In the inferred model, the three populations form three separate bands with very little overlap, although observational errors blurs this structure. Objects are classified as single/binary/trinary mainly as a function of brightness with respect to the main sequence, although neighboring data points can differ somewhat depending on their respective measurement errors. An object with an observed magnitude in the binary band can be consistent with the single population if the errors are large enough. Because the number density of the singles population is higher, such an object will be inferred to have a lower intrinsic absolute magnitude. Equivalently, an object in the trinary band with large errors will tend towards the binary population. In the model with noise, shown in the right panel, the points are given an additional scatter corresponding to the noise of the data. In order to account for a magnitude dependence to the noise, we have binned the sample region into six bins in absolute magnitude, such that the noise in the right panel is drawn at random from the corresponding bin in the data panel.

\begin{figure}[tbp]
\centering
\includegraphics[width=0.8\textwidth,origin=c]{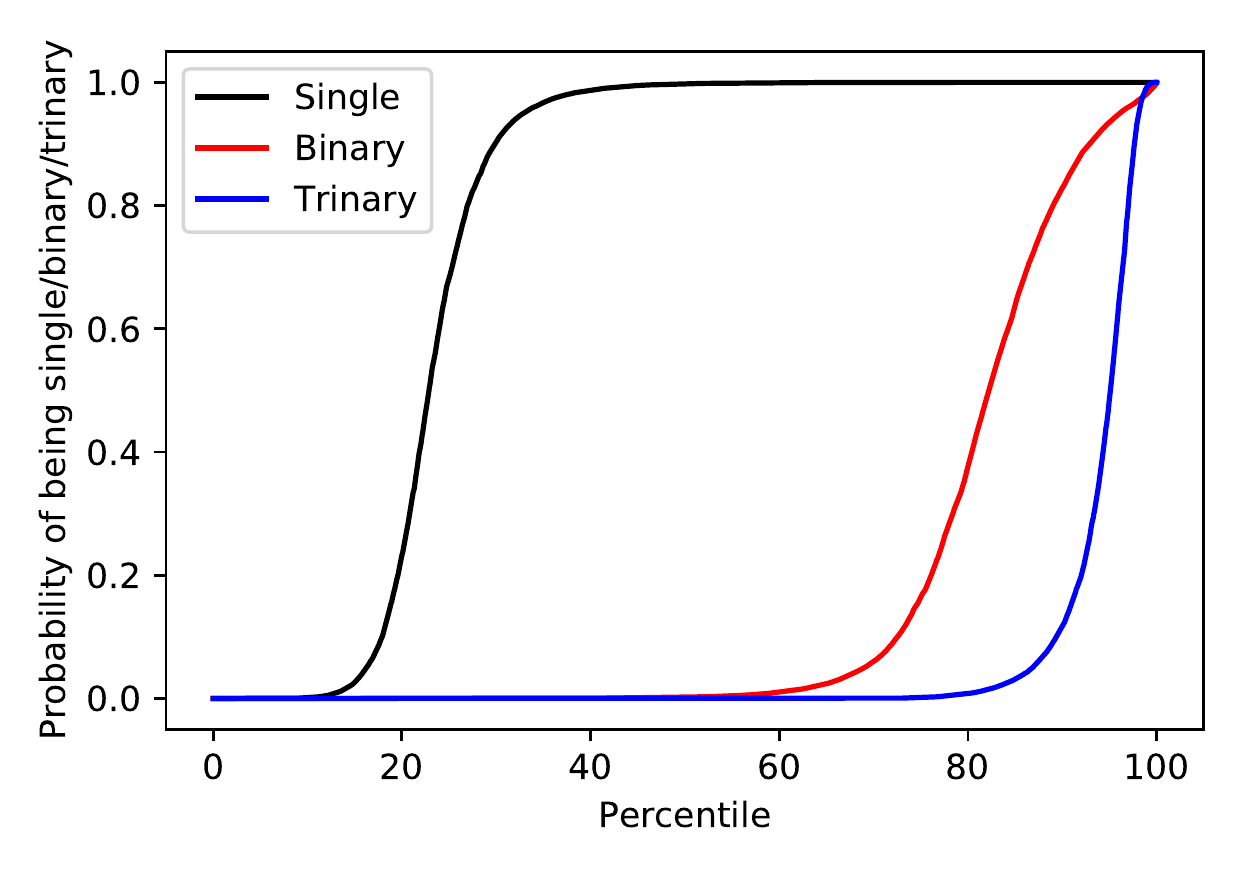}
\caption{\label{fig:percentiles_obj_type}Percentile values for the posterior probability of being single, binary, or trinary, for all stars in our sample. The posterior values are marginalized over the population parameters.}
\end{figure}

Figure \ref{fig:percentiles_obj_type} shows a summary statistic of the inferred object types. A majority of objects are very likely single stars, while strongly inferred binaries are less numerous and trinaries even more so. While very strongly inferred singles and binaries should indeed be interpreted as such, it is possible that the objects strongly inferred to be trinaries are in fact higher multiples. Including quaternary systems in this analysis would probably make an even better fit for the objects furthest from the main sequence.

\begin{figure}[tbp]
\centering
\includegraphics[width=0.7\textwidth,origin=c]{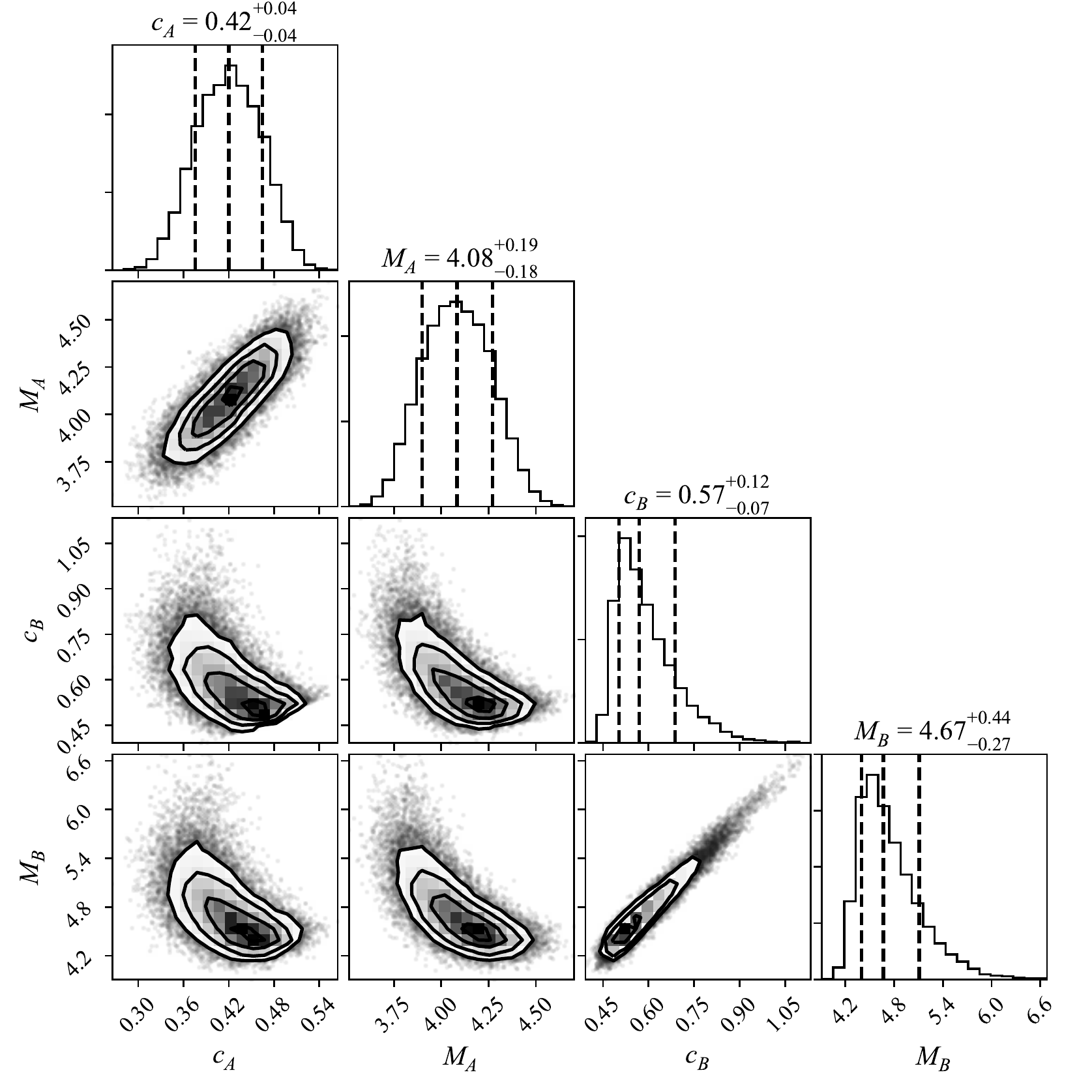}
\caption{\label{fig:disentangledbinary}Posterior values on color and absolute magnitude for the component stars of a binary system, marginalized over the population parameters. The object is the same as is used in figure \ref{fig:disentangledtrinary}.}
\end{figure}

\begin{figure}[tbp]
\centering
\includegraphics[width=\textwidth,origin=c]{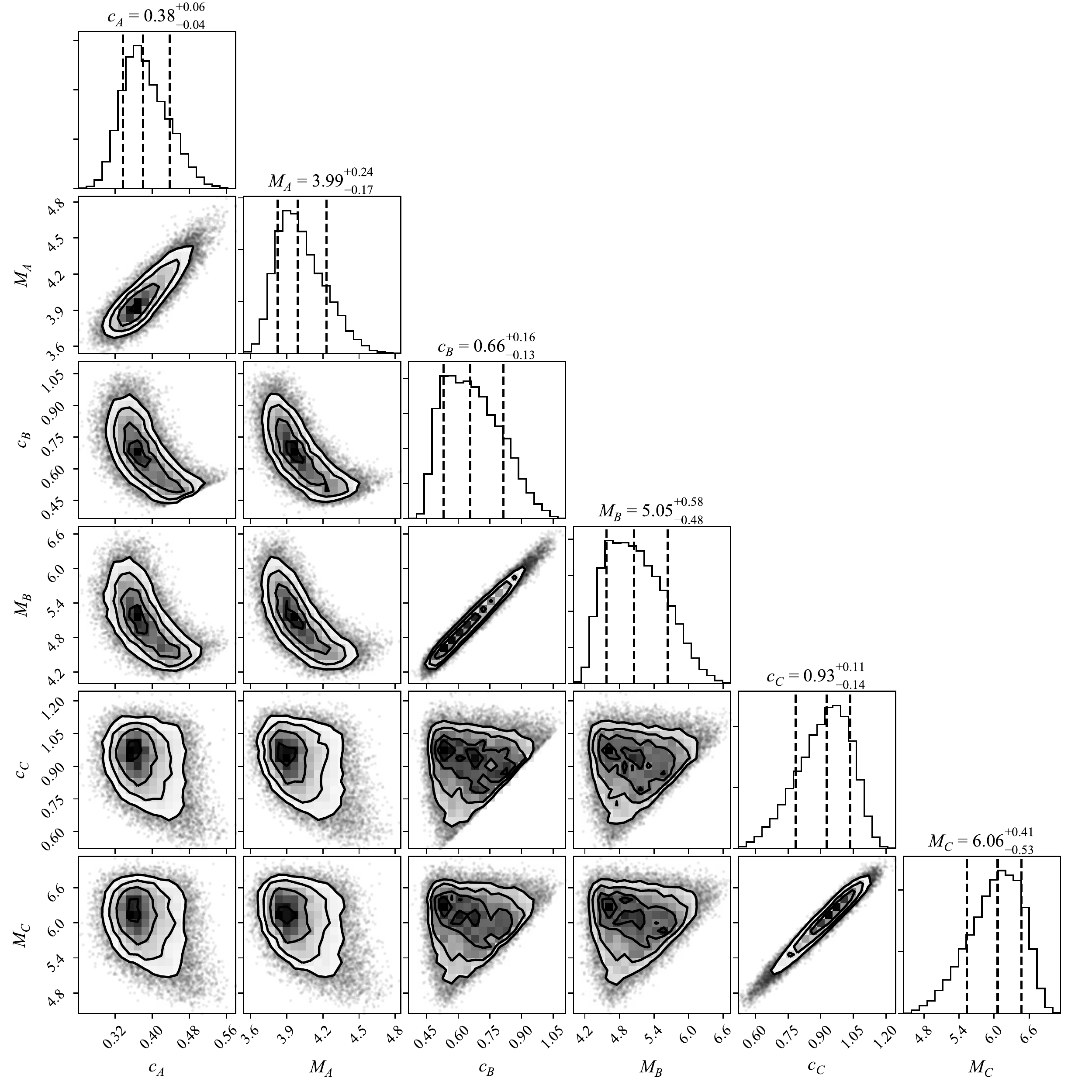}
\caption{\label{fig:disentangledtrinary}Posterior values on color and absolute magnitude for the component stars of a trinary system, marginalized over the population parameters. The object is the same as is used in figure \ref{fig:disentangledbinary}.}
\end{figure}

In figures \ref{fig:disentangledbinary} and \ref{fig:disentangledtrinary}, we see the posterior on the color and magnitude of the component stars of a multiple stellar system, where the posterior is marginalized over the population parameters. These Monte-Carlo Markov Chains were found using the \textsc{Emcee} implementation \citep{ForemanMackey2013} of the affine-invariant ensemble sampling method \citep{Goodman2010}. Both figures are for the same object, but in figure \ref{fig:disentangledbinary} it is assumed that this object is a binary system, while in figure \ref{fig:disentangledtrinary} it is assumed that it is trinary. This object\footnote{Gaia DR1 source ID: 6220743838828320384 -- 2MASS identification number: 14230246-3029168} is chosen as it has an equal probability between these two possibilities: it has a 50.0 \% probability of being binary and a 50.0 \% probability of being trinary. Its observed color and magnitude are $\hat{c}=0.512$ and $\hat{M}=3.54$. In both these figures, the posteriors are constrained by the criterion that colors are in ascending order, such that $c_A < c_B (< c_C)$. Clearly, the color and magnitude of the brighter component star is much better constrained than those of the dimmer component star(s).

\begin{figure}[tbp]
\centering
\includegraphics[width=0.8\textwidth,origin=c]{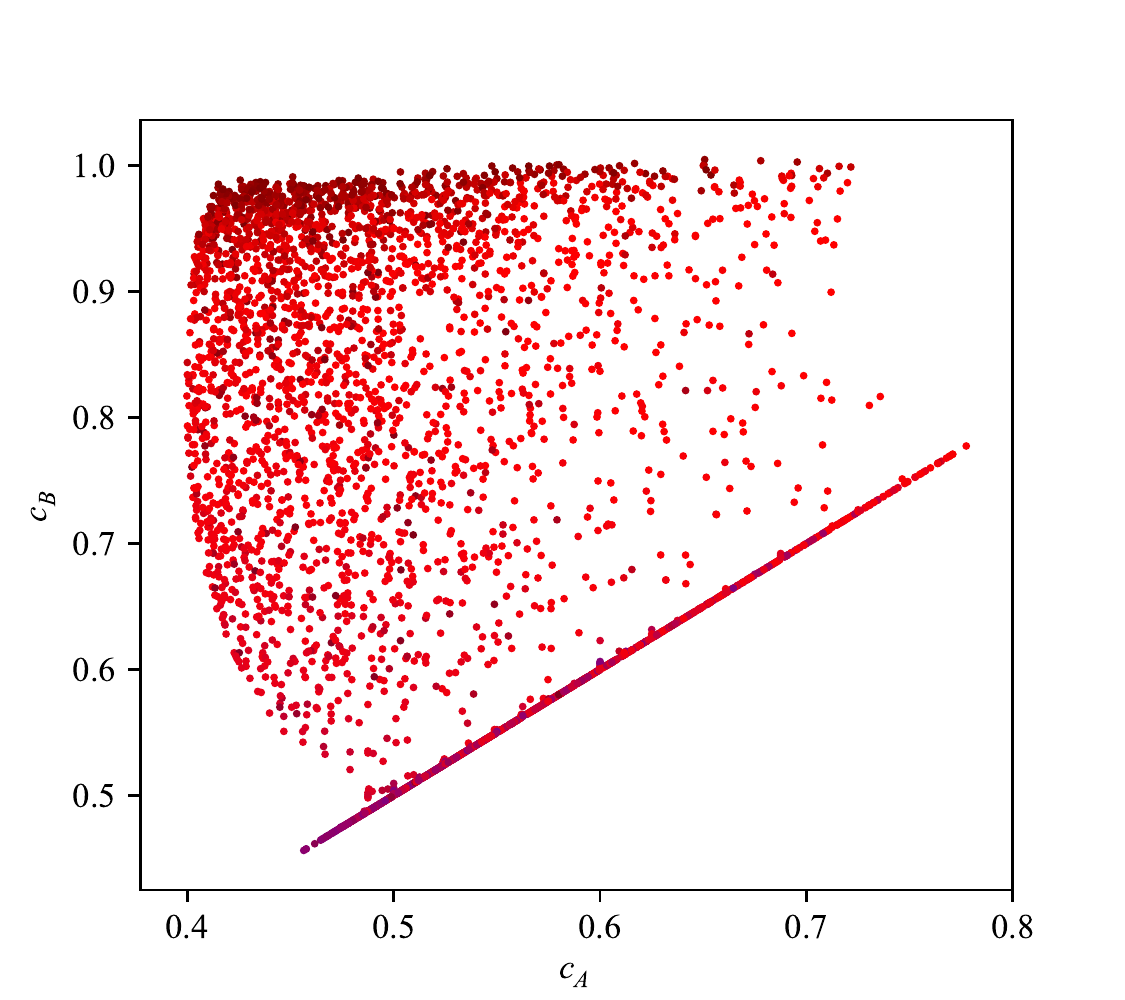}
\caption{\label{fig:binaryensemble}Maximum posterior values on the colors of the component stars of binaries, marginalized over the population parameters. Included are all stellar objects that are inferred to be binary with more than 50 per cent probability, giving a total number of 3,473 objects. The scatter points have a color code corresponding to that of figure \ref{fig:colortriangle}. The lower limit, visible as a line occupied by many objects, corresponds to the case where a binary is made from two component stars of identical color.}
\end{figure}

In figure \ref{fig:binaryensemble}, we turn to the objects of our catalog  that are inferred to be binaries with a probability greater $0.5$, and show the maximum posterior values of the component star colors for objects. It is clear that objects for which the difference in color between the component stars is large, seen towards the top of the figure, are also more consistent with being single stars, as seen by the dark red scatter points. Conversely, objects where the difference in color between the component stars is small, towards the bottom of the panel, are also more consistent with being trinaries, as seen by the purple scatter points.

\section{Discussion and conclusion}\label{sec:discussion}

We developed a flexible data-driven model of the color-magnitude diagram simultaneously fitting for populations of single stars as well as unresolved binaries and trinaries for the first time. We applied this model to photometric and astrometric data from the Gaia TGAS cross-matched with the 2MASS survey, and found that this model can explain the scatter and distribution of objects in the data. The same method can be applied to any data set where the distances to the stellar objects are known to adequate accuracy, such as an open cluster or a nearby local group galaxy.

The model itself relies on describing the population of single stars with Gaussian distributions positioned on a line, with amplitude exponentially decreasing as a function of color. 
With this model at hand, the populations of binaries and trinaries can be predicted, and we showed that despite the non-linear transformation they are also well approximated by Gaussians. Therefore, the full model is parametrized by the slope and intercept of the line supporting the Gaussians describing single stars, their covariances, the decay rate of their amplitudes as a function of color, as well as the fractional probabilities of objects being binaries or trinaries. This approach provides an accurate description of the data and is suitable for efficient inference of its parameters.

With the parameters of this model constrained from the TGAS-2MASS data, we derived probabilities for individual objects to be single, binary, or triple systems, as well as making predictions for their individual colors and magnitudes (typically strongest for the system's brightest star).
Those results are robust, with all sources of uncertainties fully included, or equivalently, all parameters marginalized over.
They minimally rely on assumptions about the underlying physics, not depending on rigid stellar models but rather on a flexible data-driven model providing a good description of the data. For simplicity, we did not include higher multiples in our analysis, as a population of singles, binaries, and trinaries describes the overall population very well. However, the systems strongly inferred to be trinary could very well be quaternary system instead. In the inferred model, visible as the center panel of figure \ref{fig:data&model}, the three populations form separate bands. With more precise parallax measurements such bands would be clearly discernible, making the inference on the model more powerful and robust.

One of the main limitations of this analysis is the absence of a model for the selection function (\textit{i.e.}, the detection and noise properties of the objects in our sample). We know for a fact that selection effects are very severe, coming from the TGAS catalogue and the complex scanning law of the Gaia survey, and also the TGAS/2MASS cross correlation. Selection effects are dependent on position on the sky, apparent magnitude, and proper motion, among other effects, and are difficult to account for. Furthermore, the chance of confusing two stars as one object is a function of distance (a binary systems might be resolved if nearby but not if far away). For those reasons, we cannot fully interpret $f_b$ and $f_t$ as actual fractions of binary and trinary systems in the solar neighborhood. Rather, they are the fraction of unresolved binary and trinary systems in our sample, where selection effects are included.

Future Gaia Data Releases will allow us to improve this model and resolve some of those limitations. 
The selection function will be simpler, and Gaia colors will be available, therefore reducing the need for external data like 2MASS.
We also intend to extend this work in the following ways: include higher multiples, model other regions of color-magnitude space, use a less restrictive cut in parallax, develop more sophisticated models of the data involving Gaussian Mixtures and stellar models. A major improvement would be to account for how the resolving power decreases with distance, which could give a more robust result and constrain the distribution of orbital separation in multiple stellar systems.

\acknowledgments

We would like to thank our referee, Maxwell Moe, for a constructive peer review that was of great benefit to this article. This work was supported by collaborative visits funded by the Cosmology and Astroparticle Student and Postdoc Exchange Network (CASPEN). BL was supported by NASA through the Einstein Postdoctoral Fellowship (award number PF6-170154). DWH was partially supported by the NSF (AST-1517237) and the Moore-Sloan Data Science Environment at NYU.

\vspace{5mm}
\facilities{
This work has made use of data from the European Space Agency (\acronym{ESA}) mission Gaia\footnote{\url{http://www.cosmos.esa.int/gaia}}, processed by the Gaia Data Processing and Analysis Consortium\footnote{\url{http://www.cosmos.esa.int/web/gaia/dpac/consortium}} (\acronym{DPAC}). Funding for the \acronym{DPAC} has been provided by national institutions, in particular the institutions participating in the Gaia Multilateral Agreement.
This publication makes use of data products from the Two Micron All Sky Survey, which is a joint project of the University of Massachusetts and the Infrared Processing and Analysis Center/California Institute of Technology, funded by the National Aeronautics and Space Administration and the National Science Foundation.
}

\software{This research utilized the following open-source Python packages: \textsc{Matplotlib} \citep{Hunter:2007}, \textsc{Emcee} \citep{ForemanMackey2013}. This work additionally used the Gaia science archive (https://gea.esac.esa.int/archive/) and the \acronym{NASA} Astrophysics Data System.}

%\appendix
\bibliography{bib}

\begin{thebibliography}{}
\expandafter\ifx\csname natexlab\endcsname\relax\def\natexlab#1{#1}\fi
\providecommand{\url}[1]{\href{#1}{#1}}

\bibitem[{{Anderson} {et~al.}(2017){Anderson}, {Hogg}, {Leistedt},
  {Price-Whelan}, \& {Bovy}}]{Anderson17}
{Anderson}, L., {Hogg}, D.~W., {Leistedt}, B., {Price-Whelan}, A.~M., \&
  {Bovy}, J. 2017, ArXiv e-prints, arXiv:1706.05055

\bibitem[{{Duch{\^e}ne} \& {Kraus}(2013)}]{Duchene13}
{Duch{\^e}ne}, G., \& {Kraus}, A. 2013, \araa, 51, 269

\bibitem[{{El-Badry} {et~al.}(2018){El-Badry}, {Rix}, {Ting}, {Weisz},
  {Bergemann}, {Cargile}, {Conroy}, \& {Eilers}}]{ElBadry17}
{El-Badry}, K., {Rix}, H.-W., {Ting}, Y.-S., {et~al.} 2018, \mnras, 473, 5043

\bibitem[{{Foreman-Mackey} {et~al.}(2013){Foreman-Mackey}, {Hogg}, {Lang}, \&
  {Goodman}}]{ForemanMackey2013}
{Foreman-Mackey}, D., {Hogg}, D.~W., {Lang}, D., \& {Goodman}, J. 2013, \pasp,
  125, 306

\bibitem[{{Gaia Collaboration} {et~al.}(2016{\natexlab{a}}){Gaia
  Collaboration}, {Prusti}, {de Bruijne}, {Brown}, {Vallenari}, {Babusiaux},
  {Bailer-Jones}, {Bastian}, {Biermann}, {Evans}, \& et~al.}]{gaia}
{Gaia Collaboration}, {Prusti}, T., {de Bruijne}, J.~H.~J., {et~al.}
  2016{\natexlab{a}}, \aap, 595, A1

\bibitem[{{Gaia Collaboration} {et~al.}(2016{\natexlab{b}}){Gaia
  Collaboration}, {Brown}, {Vallenari}, {Prusti}, {de Bruijne}, {Mignard},
  {Drimmel}, {Babusiaux}, {Bailer-Jones}, {Bastian}, \& et~al.}]{tgas}
{Gaia Collaboration}, {Brown}, A.~G.~A., {Vallenari}, A., {et~al.}
  2016{\natexlab{b}}, \aap, 595, A2

\bibitem[{{Goodman} \& {Weare}(2010)}]{Goodman2010}
{Goodman}, J., \& {Weare}, J. 2010, Communications in Applied Mathematics and
  Computational Science, Vol.~5, No.~1, p.~65-80, 2010, 5, 65

\bibitem[{{Green} {et~al.}(2015){Green}, {Schlafly}, {Finkbeiner}, {Rix},
  {Martin}, {Burgett}, {Draper}, {Flewelling}, {Hodapp}, {Kaiser}, {Kudritzki},
  {Magnier}, {Metcalfe}, {Price}, {Tonry}, \& {Wainscoat}}]{Green2015bayestar}
{Green}, G.~M., {Schlafly}, E.~F., {Finkbeiner}, D.~P., {et~al.} 2015, \apj,
  810, 25

\bibitem[{Hunter(2007)}]{Hunter:2007}
Hunter, J.~D. 2007, Computing In Science \& Engineering, 9, 90

\bibitem[{{Leistedt} \& {Hogg}(2017)}]{Leistedt17}
{Leistedt}, B., \& {Hogg}, D.~W. 2017, \aj, 154, 222

\bibitem[{{Lindegren} {et~al.}(2016){Lindegren}, {Lammers}, {Bastian},
  {Hern{\'a}ndez}, {Klioner}, {Hobbs}, {Bombrun}, {Michalik}, {Ramos-Lerate},
  {Butkevich}, {Comoretto}, {Joliet}, {Holl}, {Hutton}, {Parsons},
  {Steidelm{\"u}ller}, {Abbas}, {Altmann}, {Andrei}, {Anton}, {Bach},
  {Barache}, {Becciani}, {Berthier}, {Bianchi}, {Biermann}, {Bouquillon},
  {Bourda}, {Br{\"u}semeister}, {Bucciarelli}, {Busonero}, {Carlucci},
  {Casta{\~n}eda}, {Charlot}, {Clotet}, {Crosta}, {Davidson}, {de Felice},
  {Drimmel}, {Fabricius}, {Fienga}, {Figueras}, {Fraile}, {Gai}, {Garralda},
  {Geyer}, {Gonz{\'a}lez-Vidal}, {Guerra}, {Hambly}, {Hauser}, {Jordan},
  {Lattanzi}, {Lenhardt}, {Liao}, {L{\"o}ffler}, {McMillan}, {Mignard}, {Mora},
  {Morbidelli}, {Portell}, {Riva}, {Sarasso}, {Serraller}, {Siddiqui}, {Smart},
  {Spagna}, {Stampa}, {Steele}, {Taris}, {Torra}, {van Reeven}, {Vecchiato},
  {Zschocke}, {de Bruijne}, {Gracia}, {Raison}, {Lister}, {Marchant},
  {Messineo}, {Soffel}, {Osorio}, {de Torres}, \& {O'Mullane}}]{gaia_dr1}
{Lindegren}, L., {Lammers}, U., {Bastian}, U., {et~al.} 2016, \aap, 595, A4

\bibitem[{{Maoz} \& {Hallakoun}(2017)}]{Maoz16}
{Maoz}, D., \& {Hallakoun}, N. 2017, \mnras, 467, 1414

\bibitem[{{Moe} \& {Di Stefano}(2017)}]{Moe17}
{Moe}, M., \& {Di Stefano}, R. 2017, \apjs, 230, 15

\bibitem[{{Pecaut} \& {Mamajek}(2013)}]{Pecaut13}
{Pecaut}, M.~J., \& {Mamajek}, E.~E. 2013, \apjs, 208, 9

\bibitem[{{Smart}(2016)}]{Smart16}
{Smart}, R. 2016, {Gaia TGAS with 2MASS, WISE and Tycho2 photometry [Data
  set]},  published online.
\newblock \url{http://doi.org/10.5281/zenodo.161413}

\end{thebibliography}

%\begin{thebibliography}{}
%\bibitem[Anderson et al.(2017)]{Anderson17} Anderson, L., et al., \ 2017, arXiv:1706.05055
%\bibitem[El-Badry et al.(2017)]{ElBadry17} El-Badry, K., et al., \ 2017, arXiv:1709.03983
%\bibitem[Leistedt \& Hogg(2017)]{Leistedt17} Leistedt, B., \& Hogg, D. W., \ 2017, arXiv:1703.08112
%\bibitem[Maoz \& Hallakoun(2016)]{Maoz16} Maoz, D., \& Hallakoun, N.,\ 2016, arXiv:1609.02156
%\bibitem[Moe \& Di Stefano(2017)]{Moe17} Moe, M., \& Di Stefano, R., \ 2017, Astrophys J Suppl Ser, 230, 2, 15 
%\end{thebibliography}



\end{document}